\documentclass[12pt]{article}
\usepackage{amssymb, amsmath}

\newcommand{\NN}{\mathbb{N}}

\begin{document}

\begin{center}
{\large Comments on Beckmann's Uniform Reducts}\\
Stephen Cook\\
January, 2006
\end{center}

These comments refer to Arnold Beckmann's paper \cite{beckmann}.
That paper introduces the notion of the {\em uniform reduct} of a propositional
proof system, which consists of a collection of $\Delta_0(\alpha)$
formulas, where $\alpha$ is a unary relation symbol.  Here I will define
essentially the same thing, but make it a collection of $\Sigma^B_0$
formulas instead.  The $\Sigma^B_0$ formulas (called $\Sigma^p_0$
by Zambella) are two-sorted formulas which are the same as bounded
formulas of Peano arithmetic, except that they are allowed free
``string'' variables $X,Y,Z,...$ which range over finite sets of
natural numbers.   Terms of the form $|X|$ are allowed, which
denote the ``length'' of the string $X$ (more precisely 1 plus the
largest element of $X$, or 0 if $X$ is empty).  The atomic formula
$X(t)$ means $t$ is a member of $X$.

Each $\Sigma^B_0$ formula $\varphi(X)$ translates into a family
$\langle \varphi(X)[n]: n\in \NN \rangle$ of propositional formulas
(see \cite{survey,book})
in the style of the Paris-Wilkie translation.  The difference is
that now $X$ has a length $|X|$, and this affects the semantics of
$\varphi(X)$ and the resulting translation.  For each $n\in\NN$
the propositional translation $\varphi(X)[n]$ of $\varphi(X)$
has atoms $p^X_0,\cdots,p^X_{n-2}$ representing the bits of the string $X$,
and $\varphi(X)[n]$ is a tautology iff $\varphi(X)$ holds for all
strings $X$ of length $n$.  If $\varphi(\vec{X})$ has several string
variables $\vec{X}=X_1,\cdots,X_k$ then the translation is the
family $\varphi(\vec{X})[\vec{n}]$ of formulas, where $n_i$ is intended
to be the length of $X_i$.

In terms of $\Sigma^B_0$ formulas, the definition of uniform reduct
in \cite{beckmann} becomes

{\bf Definition:} (Beckmann) 
$$
   U_f = \{\mbox{$\varphi(\vec{X}) \in \Sigma^B_0$ :
$\langle \varphi(\vec{X})[\vec{n}]: n\in \NN \rangle$ has polysize $f$-proofs}
\}
$$

Problem 2 in \cite{beckmann} asks (in our teminology)
whether there is a proof system $f$ such that $U_f = \mbox{TRUE}_{\Sigma_0^B}$
(refering to the set of true $\Sigma_0^B$ formulas).

Here we point out that a positive answer to Problem 2 is equivalent to
the existence of an optimal proof system.

Let $f+$ be the the system
$f$ augmented to allow substitution Frege rules to be applied to tautologies
after exhibiting their $f$ proofs.

{\bf Theorem 1:} $U_{f+} = \mbox{TRUE}_{\Sigma_0^B}$ iff $f+$ simulates
every proof system.

{\bf Proof:}\\
$\Longleftarrow:$  For each $\Sigma^B_0$ formula $\varphi(\vec{X})$
we can easily define a proof system in which
$\langle \varphi(\vec{X})[\vec{n}]: n\in \NN \rangle$ has polysize proofs.

$\Longrightarrow:$  Assume $U_{f+} = \mbox{TRUE}_{\Sigma_0^B}$ and let
$g$ be any proof system.  The idea is to formulate the soundness of
$g$ as a $\Sigma^B_0$ formula $Sound_g$ and then show that EF, using
the propositional translations of $Sound_g$ as axioms, simulates $g$.

This is similar to Theorem 14.1.2 in \cite{krajicek}, which states
that $EF + \|0-RFN(g)\|$ p-simulates $g$, except soundness of $g$ is
now formulated by the formula $0-RFN(g)$, which is not $\Sigma^B_0$.
(See also \cite{pudlak}.)

To formulate soundness of $g$ by a $\Sigma^B_0$ formula we use a
$\Sigma^B_0$ formula $Eval(X,Y,Z)$ which asserts that $X$ is a
truth assignment to the atoms of the formula $Y$, and $Z$ extends
that assignment to the subformulas of $Y$ (see Definition 9.3.1.4
in \cite{krajicek}).  (The string $Z$ includes parsing information
for the formula $Y$.)

The proof system $g$ is a polynomial time map taking strings onto
the set of tautolgies.  Let $\varphi_g(U,Y,W)$ be a $\Sigma^B_0$
formula which asserts that $W$ is a computation showing that
$$
g(U) = Y
$$
Then we define
$$
Sound_g(U,W,X,Y,Z)= \qquad Eval(X,Y,Z)\wedge\varphi_g(U,Y,W) \ \supset \ Z(0)
$$
where we have rigged the formula $Eval$ so that $Z(0)$ is the truth value
of the entire formula $Y$.

If $g$ is a proof system, then the universal closure of $Sound_g$ is true,
and hence its propositional translations $Sound_g[\vec{n}]$
have polynomial size $f+$ proofs.

Now let $U_0$ be a string which is a $g$-proof of a formula $A$,
so $g(U_0)=A$.  Let $W_0$ be a computation showing $g(U_0)=A$.

Let $Sound'_g(X,Z)$ be the result of substituting $U_0,A,W_0$
for $U,Y,W$ in $Sound_g$, and simplifying $\varphi_g(U_0,A,W_0)$
to 1.  Thus
$$
Sound'_g(X,Z)=  \qquad Eval(X,A,Z) \supset Z(0)
$$
Then (for suitable $k,m$) the translation $Sound'_g(X,Z)[k,n]$
can be obtained by a short substitution Frege proof from the
tautologies $Sound_g[\vec{n}]$.  Now we continue this substitution
Frege proof by substitutions in $Sound'_g(X,Z)[k,n]$ as follows:

Substitute the atoms $q_1,...,q_\ell$ of the formula $A$ for the 
corrsponding atoms $p^X_0,\cdots , p^X_{\ell-1}$ coding the
truth assignment $X(0),\cdots, X(\ell-1)$ to $A$.

For each subformula $B$ of $A$ substitute $B$ for the corrsponding
atom $p^Z_i$, where $Z(i)$ codes the truth assignment to $B$.
In particular, substitute $A$ for $p^Z_0$.

The resulting formula has the form $Eval' \supset A$, where $Eval'$
has a short Frege proof.  Thus we obtain a $f+$ proof of $A$
which is polynomial in the length of the $g$ proof $U_0$ of $A$.
\hfill  $\Box$

{\bf Strongly Uniform Reducts}

We can strengthen the definition of uniform reduct to obtain the
notion of {\em strongly uniform reduct} of $f$ as follows:

{\bf Definition:}
$$
 SU_f = \{\mbox{$\varphi(\vec{X})\in\Sigma^B_0$ : there is a polytime function that
takes $\vec{n}$ to an $f$-proof of $\varphi(\vec{X})[\vec{n}]$}   \}
$$
where polytime means time $(\Sigma n_i)^{O(1)}$.

We can strengthen Theorem 1 for the case of strongly uniform reducts
by replacing ``simulates'' by ''p-simulates''.  If $f$ p-simulates $g$
then there is a polytime algorithm which translates $g$-proofs to
$f$-proofs, whereas if $f$ merely simulates $g$, then the poly-expanded
$f$-proof exists, but there is no guarantee it can be found in polytime.

{\bf Theorem 2:}
$SU_{f+} = \mbox{TRUE}_{\Sigma_0^B}$ iff $f+$ p-simulates
every proof system.

The proof is obtained from the proof of Theorem 1 by noticing that
we can efficiently construct the substitution Frege proofs involved
from the $f+$ proofs of $Sound_g[\vec{n}]$.  \hfill $\Box$

{\bf Remark}  As far as we know, an optimal proof system might exist
even though NP $\not=$ coNP.

\bibliographystyle{alpha}
\bibliography{reducts}

\end{document}